\documentclass[twocolumn]{aastex61}
\usepackage{ragged2e}

\newcommand{\pisco}{J1342+0928}

\newcommand{\kms}{{\rm km\,s}\ensuremath{^{-1}}}

\newcommand{\mgii}{Mg\ensuremath{\,\textsc{ii}}}
\newcommand{\cii}{[C\ensuremath{\,\textsc{ii}}]}

\newcommand{\civ}{C\ensuremath{\,\textsc{iv}}}

\accepted{for publication in ApJ Letters, March 6, 2018}

\shorttitle{X-rays from a $z=7.54$ quasar}
\shortauthors{Ba\~nados et al.}

\begin{document}

\title{\textit{Chandra} X-rays from the redshift 7.54 quasar ULAS~J1342+0928}

\correspondingauthor{Eduardo Ba\~nados}
\email{ebanados@carnegiescience.edu}

\author[0000-0002-2931-7824]{Eduardo Ba\~nados}
\altaffiliation{Carnegie-Princeton Fellow}
\affiliation{The Observatories of the Carnegie Institution for Science, 813 Santa Barbara St., Pasadena, CA 91101, USA}

\author[0000-0002-7898-7664]{Thomas Connor}
\affiliation{The Observatories of the Carnegie Institution for Science, 813 Santa Barbara St., Pasadena, CA 91101, USA}

\author{Daniel Stern}
\affiliation{Jet Propulsion Laboratory, California Institute of Technology, 4800 Oak Grove Drive, Pasadena, California
91109, USA}

\author{John Mulchaey}
\affiliation{The Observatories of the Carnegie Institution for Science, 813 Santa Barbara St., Pasadena, CA 91101, USA}

\author{Xiaohui Fan}
\affiliation{Steward Observatory, The University of Arizona, 933 North Cherry Avenue, Tucson, AZ 85721--0065, USA}

\author{Roberto Decarli}
\affiliation{INAF -- Osservatorio di Astrofisica e Scienza dello Spazio, via Gobetti 93/3, 40129, Bologna, Italy}

\author{Emanuele~P.~Farina}
\affiliation{Department of Physics, Broida Hall, University of California, Santa Barbara, CA 93106--9530, USA}

\author{Chiara Mazzucchelli}
\affiliation{{Max Planck Institut f\"ur Astronomie, K\"onigstuhl 17, D-69117, Heidelberg, Germany}}

\author[0000-0001-9024-8322]{Bram P. Venemans}
\affiliation{{Max Planck Institut f\"ur Astronomie, K\"onigstuhl 17, D-69117, Heidelberg, Germany}}

\author{Fabian Walter}
\affiliation{{Max Planck Institut f\"ur Astronomie, K\"onigstuhl 17, D-69117, Heidelberg, Germany}}

\author{Feige Wang}
\affiliation{Department of Physics, Broida Hall, University of California, Santa Barbara, CA 93106--9530, USA}

\author{Jinyi~Yang}
\affiliation{Steward Observatory, The University of Arizona, 933 North Cherry Avenue, Tucson, AZ 85721--0065, USA}

\begin{abstract}
We present a 45\,ks \textit{Chandra} observation of the quasar ULAS~J1342+0928 at $z=7.54$. We detect $14.0^{+4.8}_{-3.7}$ counts from the quasar in the observed-frame energy range 0.5--7.0 keV (6$\sigma$ detection), representing the most distant non-transient astronomical source identified in X-rays to date.
The present data are sufficient only to infer rough constraints on the spectral parameters. We find an X-ray hardness ratio of $\mathcal{HR} = -0.51^{+0.26}_{-0.28}$ between the 0.5--2.0 keV and 2.0--7.0 keV ranges and derive a power-law photon index of $\Gamma = 1.95^{+0.55}_{-0.53}$.
Assuming a typical value for high-redshift quasars of $\Gamma = 1.9$, ULAS~J1342+0928 has a 2--10 keV rest-frame X-ray luminosity of $L_{2-10} = 11.6^{+4.3}_{-3.5} \times 10^{44}\ {\rm erg}\ {\rm s}^{-1}$. Its X-ray-to-optical power-law slope is $\alpha_{\rm OX}=-1.67^{+0.16}_{-0.10}$, consistent with the general trend indicating that the X-ray emission in the most bolometrically powerful  quasars is weaker relative to their optical emission.
\end{abstract}

\keywords{cosmology: observations --- cosmology: early universe
--- quasars: individual (ULAS J134208.10+092838.61)}



\section{Introduction} \label{sec:intro}
Supermassive black holes are thought to be ubiquitous in the centers of massive galaxies, but their formation mechanism is still an outstanding question in astrophysics. The existence of distant quasars powered by $\gtrsim 10^9\,M_\odot$ black holes within the first Gyr of the universe sets one of the strongest challenges for supermassive black hole formation theories \citep[e.g., ][]{volonteri2012a}. X-ray observations provide a unique tool to explore the immediate vicinities of the central black holes in active galactic nuclei (AGNs; e.g., \citealt{fabian2016}). Studying the evolution of X-ray properties and trends with luminosity across cosmic time teaches us about the physics of the inner regions of AGNs, potentially providing clues about the formation of supermassive black holes.

More than 200 quasars are now known within the first Gyr of the universe \citep[i.e., at $z>5.5$; e.g.,][]{banados2016, wang-feige2016} but only about a dozen of them have been robustly detected in X-rays (i.e., $\gtrsim 5$ photons), only one of which is at $z>6.5$  \citep{moretti2014,page2014}. Pioneering X-ray studies of the $z\sim 6$ quasar population based on $5-15$\,ks \textit{Chandra} observations showed that their average X-ray properties are similar to those of low-redshift luminous quasars \citep{brandt2002,shemmer2006}. Recent deeper ($\gtrsim 50\,$ks) \textit{Chandra} and \textit{XMM-Newton} observations of a few $z>6$ quasars have permitted meaningful constraints on key parameters of individual sources such as their photon index  \citep[$\Gamma$; e.g.,][]{ai2017,gallerani2017b,nanni2018}, which is a tracer of the black hole accretion rate \citep[e.g.,][]{brightman2013}.

Here, we report \textit{Chandra} observations of the recently discovered redshift-record quasar ULAS~J134208.10 + 092838.61 (hereafter \pisco) at $z=7.54$ \citep{banados2018}. This quasar has a $7.8^{+3.3}_{-1.9}\times 10^8\,M_\odot$ accreting black hole 
and resides in a massive galaxy that is intensively forming stars \citep[$85-545\,M_\odot\,\mathrm{yr}^{-1}$;][]{venemans2017c}. Based on a $2.6\,\sigma$ detection at 41\,GHz, \cite{venemans2017c} classify this quasar as being potentially radio-loud with $R=S_{5,\,\mathrm{GHz, rest}}/S_{4400,\,\mathrm{\AA, rest}} = 12.4$ (see discussion in their Section 3.2). Deeper radio observations are already accepted to confirm this preliminary result.

To directly compare with results from the literature \citep[e.g.,][]{nanni2017}, we use a flat cosmology
with $H_0 = 70 \,\mbox{km\,s}^{-1}$\,Mpc$^{-1}$, $\Omega_M = 0.3$, and $\Omega_\Lambda = 0.7$. We assume a Galactic absorption column density toward \pisco\ of $N_{\rm H} = 1.61 \times 10^{20}\ {\rm cm}^{-2}$ \citep{2005A&A...440..775K}. Errors are reported at the 1$\sigma$ (68\%) confidence level unless otherwise stated. Upper limits correspond to $3\sigma$ limits.

\section{Observations and Data Reduction}
\begin{figure}
\plotone{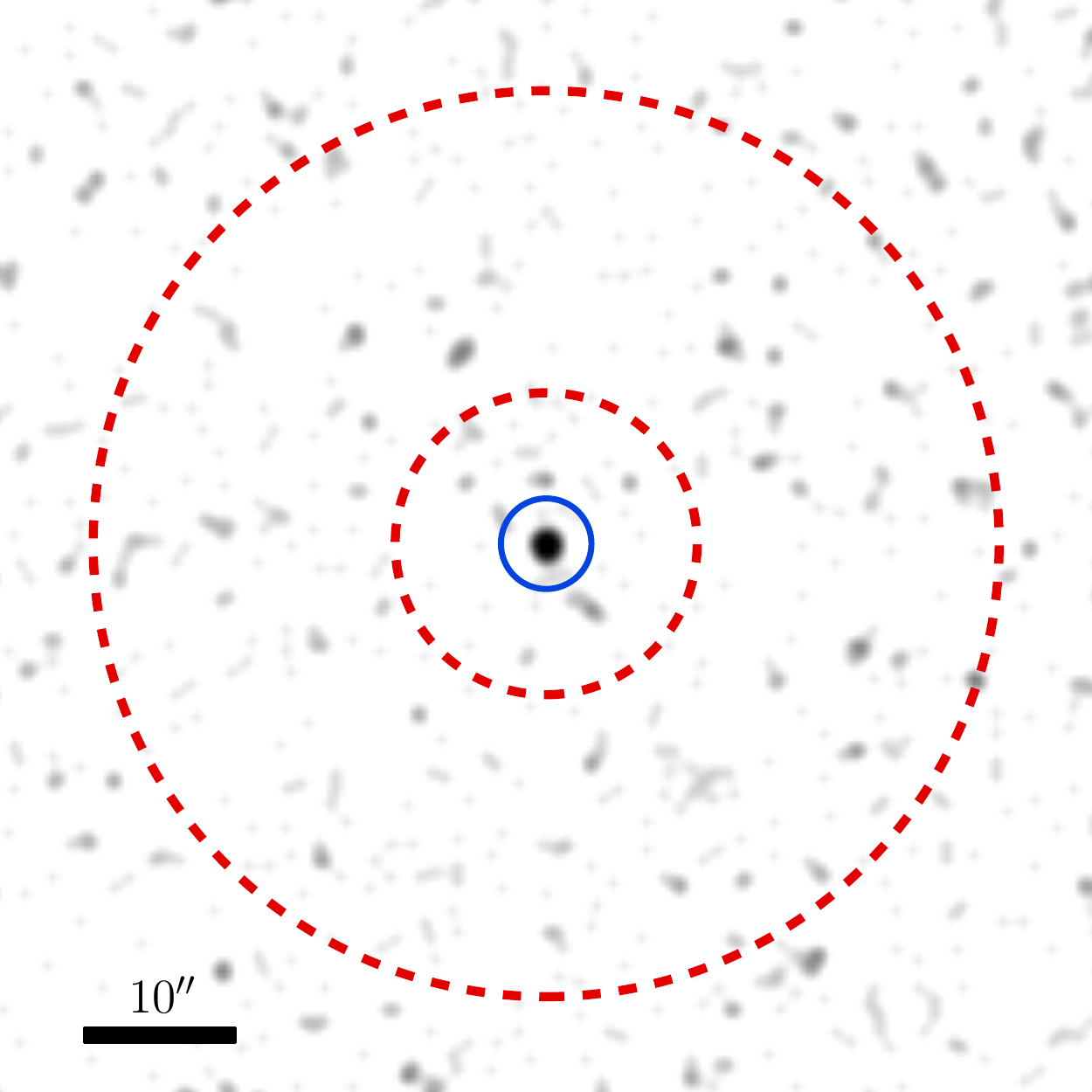}
\caption{{\it Chandra} observation of \pisco. This image covers the energy range of 0.5--7.0\,keV, was binned to pixels of size $0\farcs49$, and has been smoothed with a Gaussian kernel of width 1.5 pixels ($\sim$$0\farcs75$). The spectral extraction regions for the source (blue circle) and background (red annulus) are shown. We detect an X-ray source at the $6\sigma$ level within 1 pixel of the position reported by \cite{banados2018}. \label{fig:xray_image}}
\end{figure}

We observed \pisco\ with {\it Chandra} for a total of 45.1 ks on 2017 December 15 (24.7 ks, Obs ID: 20124) and 2017 December 17 (20.4 ks, Obs ID: 20887). The detection image is shown in Figure \ref{fig:xray_image}. Observations were conducted with the Advanced CCD Imaging Spectrometer \citep[ACIS;][]{2003SPIE.4851...28G} using the Very Faint telemetry format and the Timed Exposure mode. {\it Chandra} was pointed so that \pisco\ fell on the ACIS-S3 chip. We analyzed these data using {\tt CIAO} version 4.9 \citep{2006SPIE.6270E..1VF} and CALDB version 4.7.7. Observations were combined using the {\tt MERGE\_OBS} routine to create images in the broad (0.5--7.0 keV) energy band.  We used the ACIS standard filters for event grades (0, 2, 3, 4, and 6) and good time intervals.

We performed a source detection using {\tt WAVDETECT} \citep{2002ApJS..138..185F}, using scales of 1, 2, 4, 8, and 16 pixels and a point-spread function (PSF) map made from the weighted means of the two observations. A source was detected within $0\farcs19$ of \pisco, with $14.0^{+4.8}_{-3.7}$ net counts and a detection significance (net counts divided by background error) of $6\sigma$. Due to the low number of counts in this source, we use the \citet{gehrels1986} approximation for uncertainties. {\tt WAVDETECT} reported a PSFRATIO of 0.36 for this object, meaning that we do not find extended structure around this source and thus no evidence of powerful X-ray jets at our sensitivity (see also \citealt{fabian2014}). No other sources were detected within $1\farcm0$ of \pisco.

 To characterize this source, we used {\tt dmextract} to extract the total counts from 0.5--7.0 keV in a $3\farcs0$ radius source aperture and a $10\farcs0$ to $30\farcs0$ background annulus. We detect $13.7^{+5.1}_{-4.0}$ net counts (in agreement with the {\tt WAVDETECT} measurement), with an expected background of $2.3$ counts. Following the binomial analysis technique used by \citet{2014ApJ...785...17L}, we compute a probability of the X-ray detection of \pisco\ being a false positive: $P = 4.3 \times 10^{-9}\ (5.8\sigma$ detection). Using the Bayesian techniques described by \cite{2006ApJ...652..610P}, assuming uniform (Jeffreys) priors\footnote{Assuming flat priors on $\theta$ only changed our reported value of $\mathcal{HR}$ by ${\sim}0.25\sigma$.} and integrating the joint posterior distribution with Gaussian quadrature in 1,000 bins, we compute an X-ray hardness ratio\footnote{$\mathcal{HR} = (H-S)/(H+S)$, where $H$ and $S$ are the net counts in the hard (2.0--7.0 keV) and soft (0.5--2.0 keV) bands, respectively.} of $\mathcal{HR} = -0.51^{+0.26}_{-0.28}$ across the 0.5--2.0 keV and 2.0--7.0 keV bands.

We extracted spectra in a $3\farcs0$ radius aperture and a $10\farcs0$ to $30\farcs0$ background annulus from both observations around \pisco. Spectra were extracted in the energy range of  0.5--7.0 keV (4.3--59.8 keV in the rest frame of the quasar). Due to slight variations between the two pointings, an accurate response file cannot be generated from the combined image, so we instead extracted spectra from the two observations and combined them with the {\tt CIAO} task {\tt COMBINE\_SPECTRA}. Fitting was performed with {\tt PyXspec}, using {\tt XSPEC} version 12.9.1  \citep{1996ASPC..101...17A}.

\begin{figure}
\plotone{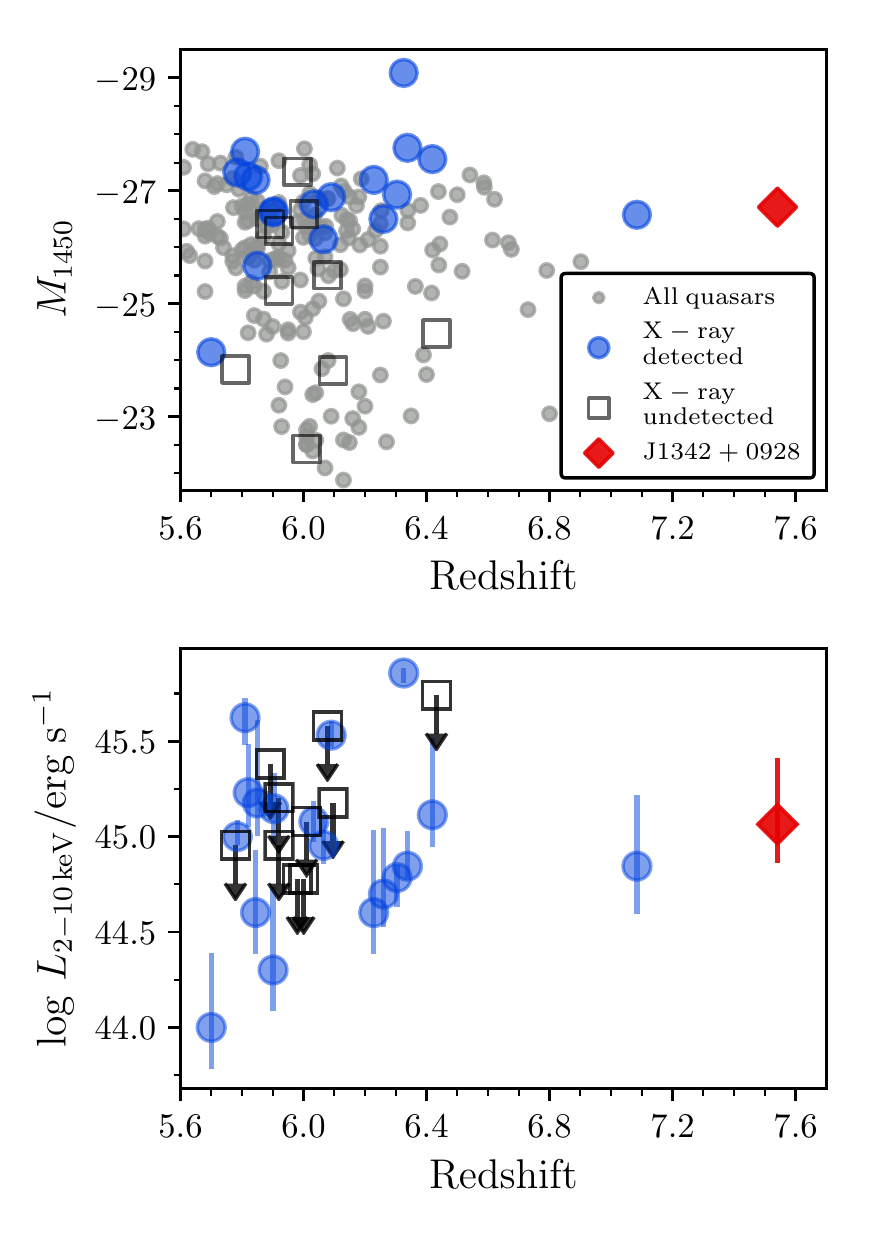}
\caption{Top: absolute magnitude at rest-frame wavelength $1450\,$\AA\ for all $z>5.6$ quasars known to date. High-redshift quasars with X-ray observations (as compiled by \citealt{nanni2017,nanni2018}) are represented by filled blue circles for detections and open squares for non-detections. The quasar studied in this work, \pisco, is shown as a red diamond.  Bottom: rest-frame $2-10$\,keV luminosity for $z>5.6$ quasars with X-ray observations\textsuperscript{$a$}. Symbols follow the same legend as the top panel. The uncertainties in the X-ray luminosity for \pisco\ correspond to the values allowed by assuming a photon index $\Gamma=1.6$ and $\Gamma=2.2$, while the data point is shown assuming our fiducial value of $\Gamma=1.9$. Upper limits are reported at the $3\sigma$ level.
\label{fig:redshift_behavior}}
\justify
\noindent {\footnotesize{\textsuperscript{$a$} All X-ray luminosities are taken from \cite{nanni2017} except for \pisco\ (this work) and J1030+0524 at $z=6.31$, which is taken from \cite{nanni2018} who reported an updated value of the rest-frame $2-8$\,keV luminosity.}}
\end{figure}

We modeled the quasar spectrum with a combined power-law model (powerlaw) with Galactic absorption (phabs),  using the modified c-statistic \citep{1979ApJ...228..939C, 1979ApJ...230..274W} to find best-fits and uncertainties. The source redshift was frozen to $z = 7.5413$ \citep{venemans2017c} and the Galactic column density to $N_{\rm H} = 1.61 \times 10^{20}\ {\rm cm}^{-2}$ \citep{2005A&A...440..775K}. When allowed to vary, we found a best-fit value of $\Gamma = 1.95^{+0.55}_{-0.53}$. In a similar situation, \cite{ai2016} reported a slope of $\Gamma = {3.0}_{-0.7}^{+0.8}$ with a 14-count {\it Chandra} spectrum of a quasar at redshift $z=6.33$. However, this value was then updated by \cite{ai2017} to $\Gamma = 2.3 \pm 0.1$ using a deeper \textit{XMM-Newton} spectrum with ${\sim}$$460$ counts.
Due to the low number of source counts in our \textit{Chandra} observation, we therefore  base our fiducial values on a fit where we froze the power-law photon index to $\Gamma=1.9$; this is choice is based on the mean power-law index found for 10 $z\sim6$ quasars studied by \citet{nanni2017}.
The only free parameter was therefore the normalization of the power law. While \cite{venemans2017c} report a dust-rich host galaxy, its absorption effect is minor given the high energies at the redshift of the quasar we are probing with the \textit{Chandra} observation.  Even if we assume an absorber at the redshift of the X-ray source with column density $N_{\rm H} = 10^{23.5}\ {\rm cm}^{-2}$, the change in our measured luminosity compared to the fiducial value is less than the uncertainties in the original measurement. As higher levels of obscuration for sources with comparable X-ray luminosity is only expected in dust-obscured galaxies (DOGS) and hot DOGS \citep{vito2018b}, we cannot place any significant constraints on this absorption without deeper observations.

Assuming a power-law index of $\Gamma = 1.9$, we find that the absorption-corrected flux of \pisco\ from 0.5--2.0 keV (observed) is $F_{0.5-2.0} = 1.68^{+0.52}_{-0.44}  \times 10^{-15}\ {\rm erg}\ {\rm s}^{-1}\ {\rm cm}^{-2}$ and the luminosity at rest-frame 2--10\,keV is $L_{2-10} = 11.6^{+4.3}_{-3.5} \times 10^{44}\ {\rm erg}\ {\rm s}^{-1}$. As the spectral response of our {\it Chandra} observations did not cover the full 2--10\,keV rest-frame energy band, the measured luminosity is extrapolated using the best-fit model. To account for the uncertainty in the value of $\Gamma$, we also find the flux and luminosities for models with $\Gamma = 1.6$ and $\Gamma = 2.2$.
We find for $\Gamma = 1.6$,
$F_{0.5-2.0} = 1.44^{+0.45}_{-0.37}  \times 10^{-15}\ {\rm erg}\ {\rm s}^{-1}\ {\rm cm}^{-2}$ and
$L_{2-10} = 8.9^{+3.3}_{-2.7}  \times 10^{44}\ {\rm erg}\ {\rm s}^{-1}$,
while for $\Gamma = 2.2$,
$F_{0.5-2.0} = 1.94^{+0.60}_{-0.50}  \times 10^{-15}\ {\rm erg}\ {\rm s}^{-1}\ {\rm cm}^{-2}$ and
$L_{2-10} = 15.3^{+5.7}_{-4.6}  \times 10^{44}\ {\rm erg}\ {\rm s}^{-1}$.
When using the free-to-vary value of $\Gamma$ ($\Gamma = 1.95$), we measure $F_{0.5-2.0} = 1.71^{+0.54}_{-0.45}  \times 10^{-15}\ {\rm erg}\ {\rm s}^{-1}\ {\rm cm}^{-2}$ and
$L_{2-10} = 13.0^{+4.0}_{-3.4}  \times 10^{44}\ {\rm erg}\ {\rm s}^{-1}$.

\section{Discussion}

\begin{figure}
\plotone{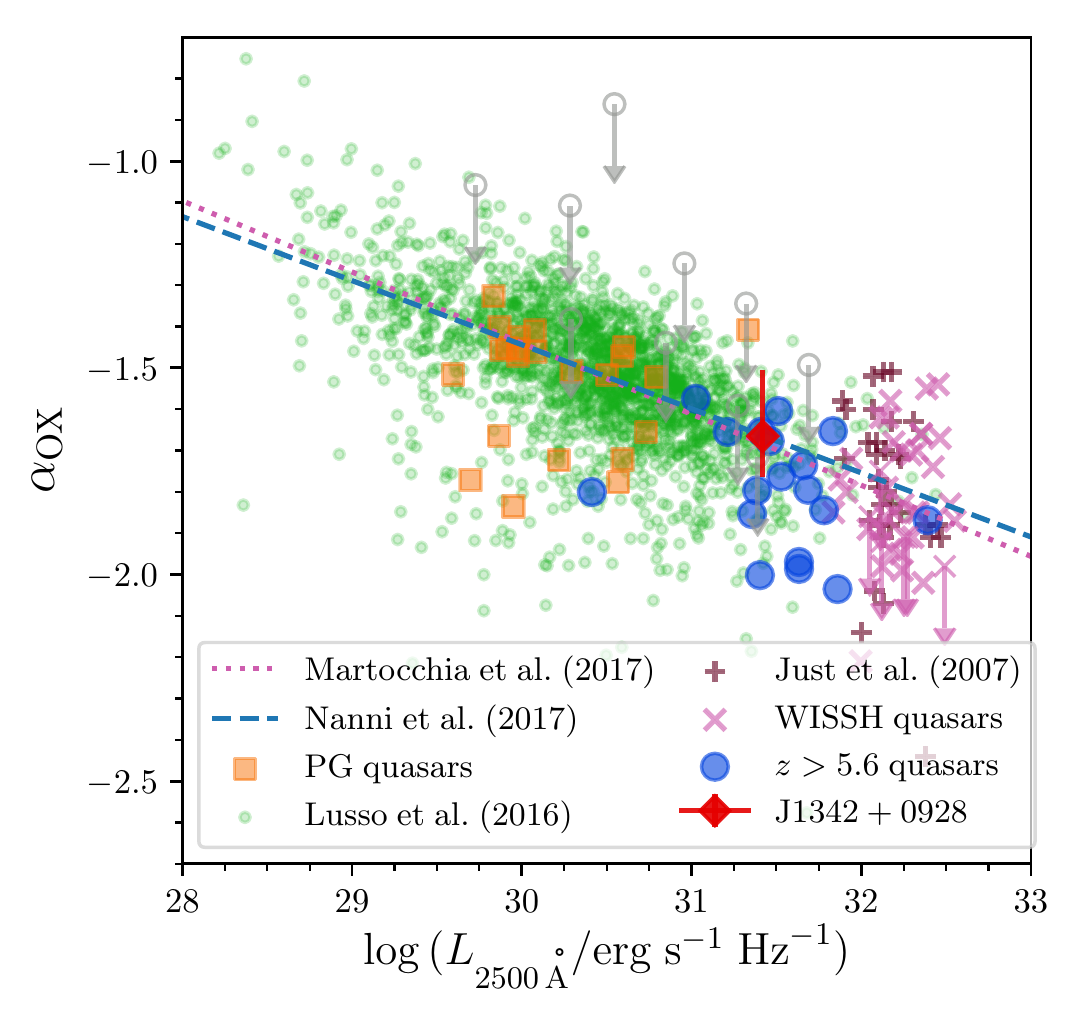}
\caption{X-ray-to-Optical power law slope ($\alpha_{\rm OX}$) as a function of the rest-frame $2500$\,\AA\ monochromatic luminosity. The green small circles represent the AGNs from \cite{lusso2016} with X-ray detections with signal-to-noise ratio greater than five.  We also show the quasars from the Palomar-Green (PG) bright quasar survey \citep{laor1994}, the hyperluminous SDSS quasars studied by \cite{just2007}, and the \textit{WISE}/SDSS selected hyperluminous (WISSH) quasar sample \citep{martocchia2017} as orange rectangles, maroon plusses, and pink crosses, respectively. The $z>5.6$ quasars (circles) and \pisco\ (red diamond; this work) occupy the locus in between the \cite{lusso2016} AGNs and the quasar samples of \cite{just2007} and \cite{martocchia2017}. In the calculation of $\alpha_{\rm OX}$ for the $z>5.6$ quasars we used the X-ray fluxes reported by \cite{nanni2017,nanni2018} and the rest-frame 2500\,\AA\ flux densities, extrapolating the 1450\,\AA\ rest-frame magnitudes (see Figure \ref{fig:redshift_behavior}), assuming a typical ultraviolet-optical power-law slope of $\alpha_\nu=-0.5$ \citep[e.g.,][]{banados2015a}. The dotted and dashed lines are the best-fit relation between $\alpha_{\rm OX}$ and $L_{2500\,{\rm \AA}}$ reported by \cite{martocchia2017} and \cite{nanni2017}, respectively. The uncertainty in $\alpha_{\rm OX}$ for \pisco\ is dominated by the systematic uncertainty in the X-ray luminosity as it was calculated fixing the power-law photon index to $\Gamma=1.9$ (see text).
\label{fig:alpha-ox}}
\end{figure}

The X-ray detection of \pisco\ at $z=7.54$ is robust (see Figure \ref{fig:xray_image}), but deeper \textit{Chandra} or \textit{XMM-Newton} observations will be crucial in order to derive meaningful X-ray spectral parameters. \pisco\ represents the most distant  non-transient
astronomical source identified in X-rays to date\footnote{We note that gamma-ray bursts have been detected in X-rays up to $z= 8.23^{+0.07}_{-0.08}$ \citep{tanvir2009}. See \cite{salvaterra2015} for a summary of the highest-redshift gamma-ray bursts.} and its $L_{2-10} = 11.6^{+4.3}_{-3.5} \times 10^{44}\ {\rm erg}\ {\rm s}^{-1}$ luminosity is consistent with the luminosities observed in other $z>5.6$ quasars (see Figure \ref{fig:redshift_behavior}).

The properties of the most distant quasars are so extreme that it is often challenging to find lower-redshift quasars with analogous properties (see, e.g., discussion in \citealt{banados2018,davies2018}).
Recently, in a series of papers, \cite{bischetti2017}, \cite{duras2017}, \cite{martocchia2017}, and \cite{vietri2018} reported the physical properties of the $z\sim 2 - 4$  \textit{WISE}/SDSS selected hyperluminous (WISSH) quasar sample. The WISSH quasars have bolometric luminosities $L_{\rm Bol} \gtrsim 10^{47}\,{\rm erg}\ {\rm s}^{-1}$, $\gtrsim 10^9\,  M_\odot$ black holes accreting near the Eddington limit, host galaxies with star formation rates up to $\sim$$2000\,M_\odot\,$yr$^{-1}$, and  show signatures of powerful  ionized outflows manifested as broad [O\,{\sc iii}] emission lines and C\,{\sc iv} blueshifts greater than $2000\,$km\,s$^{-1}$. These characteristics are similar and in some cases even more extreme than the properties of the most distant quasars known to date \citep[cf. ][]{leipski2014,mazzucchelli2017b}.

\cite{martocchia2017} reported that the WISSH quasars show a low X-ray-to-optical flux ratio compared with other lower-redshift AGNs/quasars with similar X-ray luminosities (see their Figure 4). A similar trend had already been reported for some of the most optically hyperluminous ($M_i < -29$) SDSS quasars studied by \cite{just2007} at $z\sim 1.5 - 4.5$. Other extreme sources such as the hyperluminous $z\sim 2$ hot DOGS studied by \cite{vito2018b} also show hints of weaker X-ray emission than expected from their large mid-infrared luminosities. These studies are suggesting that the X-ray flux saturates for more bolometrically powerful quasars, while the ultraviolet and mid-infrared luminosities still increase \citep[e.g.,][]{stern2015,chen_c2017}, although larger samples of the most luminous objects are still required to robustly confirm this picture.
The lamppost model of the X-ray corona (e.g., \citealt{miniutti2004})
provides a plausible physical scenario to explain this observation.  In this model, the X-ray corona is a compact plasma along the black hole spin axis, potentially associated with the base of the jet.  The hot corona Compton upscatters photons from the accretion disk to higher energies, creating much of the high-energy emission observed from AGNs. In this lamppost model, more luminous systems might push the corona to greater heights, thereby lowering its geometrical cross section with the seed accretion disk photons, and causing the observed sublinear relation between X-ray emission and both ultraviolet and mid-infrared emission.

Interestingly, intrinsically weak X-ray emission has also been
interpreted as a requirement to explain powerful quasar outflows in order to avoid the suppression of ultraviolet line driven winds \citep{richards2011,luo2013,luo2015}. \cite{mazzucchelli2017b} recently showed that large \civ\ blueshifts are common in quasars at $z>6.5$ (with a median of $\sim$$2400\,\kms$).  \pisco\ is one of the most extreme examples among these objects, presenting a \civ\ emission line blueshifted by $\sim 6100$\,km\,s$^{-1}$ from the Mg\,{\sc ii} line and $\sim 6600$\,km\,s$^{-1}$ from the systemic redshift based on the \cii\ emission line \citep{banados2018}. This gives us the opportunity to test whether these powerful quasars with strong outflows at very different redshifts have common X-ray properties.

The X-ray-to-optical flux ratio $(L_{2-10}/L_{2500{\mathrm{ \AA}}})$ for \pisco\ is $0.04$, which is similar to the values reported for the WISSH sample (see Figure 4 in \citealt{martocchia2017}). A more common term used in the literature is the X-ray-to-optical power-law slope defined as the index of the power law connecting
 the monochromatic luminosities at $2\,$kev and $2500\,$\AA: $\alpha_{\rm OX}= 0.3838 \times \log(L_{2\, \mathrm{keV}}/L_{2500\,\mathrm{\AA}})$.
In Figure \ref{fig:alpha-ox} we show $\alpha_{\rm OX}$ as a function of the rest-frame $2500\,$\AA\ monochromatic luminosity for all $z\gtrsim 6$ quasars compiled by \cite{nanni2017,nanni2018}, \pisco\ (this work), and the \citeauthor{just2007} and \citeauthor{martocchia2017} quasar samples. With $\alpha_{\rm OX}=-1.67^{+0.16}_{-0.10}$ and $L_{2500\,{\mathrm{\AA}}}=10^{31.42}\,\mathrm{erg}\,\mathrm{s}^{-1}\,\mathrm{Hz}^{-1}$, measured directly from its spectrum, \pisco\ falls right on the relations found by \cite{martocchia2017} and \cite{nanni2017} (see Figure \ref{fig:alpha-ox}). We note that the reported $\alpha_{\rm OX}$ assumes that the quasar luminosity did not vary between the near-infrared and X-ray observations, which were taken about nine months apart.  Almost all of the most distant quasars occupy a locus in Figure \ref{fig:alpha-ox} that bridges the typical AGN \citep{lusso2016} and the extreme WISSH quasars and the hyperluminous quasars from \cite{just2007}.
The lone exception is RD~J1148+5253 at $z=5.7$ \citep{mahabal2005}, which is one of the optically faintest quasars known at $z>5.5$. The properties of RD~J1148+5253 are more similar to the quasars from thee Palomar-Green (PG) bright quasar survey \citep{laor1994}. The WISSH and the hyperluminous SDSS quasars therefore seem to share similar X-ray and optical properties with the most distant quasars. A detailed comparison between these two samples could shed light into the understanding of some of the most extreme sources known in the universe.

Recently, \cite{risaliti2015} and \cite{lusso2017} have proposed that under some assumptions quasars can be used as standard candles. This is based on the tight relation between the X-ray and ultraviolet emission (see Figure \ref{fig:alpha-ox}) coupled with the full-width at half maximum (FWHM) of the \mgii\ line.
At face value, \pisco\ follows the relation ($L_{2\rm{keV}}\propto L_{2500\rm{\AA}}^{4/7}\,v_{\rm{fhwm}}^{4/7}$) reported in Figure 2 of \cite{lusso2017} within the uncertainties. Although this will need to be confirmed with deeper data and more $z>7$ quasars, the present X-ray detection of an accreting black hole at $z=7.54$  opens up the exciting prospect of testing the cosmological model when the universe was only 5\% of its current age.

\acknowledgments
We thank the referee for useful comments that improved this Letter.
 We thank Silvia Martocchia for providing the data to produce Figure \ref{fig:alpha-ox}. The work of D.S. was carried out at the Jet Propulsion Laboratory, California Institute of Technology, under a contract with NASA. B.P.V.\ and F.Walter\ acknowledge funding through ERC grants ``Cosmic Dawn" and ``Cosmic Gas".
The scientific results reported in this article are based on observations made by the \textit{Chandra} X-ray Observatory. This research has made use of software provided by the \textit{Chandra} X-ray Center (CXC) in the application package CIAO.

\vspace{5mm}

\facility{CXO}


\software{Astropy \citep{astropy2018}}


\begin{thebibliography}{}
\expandafter\ifx\csname natexlab\endcsname\relax\def\natexlab#1{#1}\fi
\providecommand{\url}[1]{\href{#1}{#1}}

\bibitem[{{Ai} {et~al.}(2016){Ai}, {Dou}, {Fan}, {Wang}, {Wu}, \&
  {Bian}}]{ai2016}
{Ai}, Y., {Dou}, L., {Fan}, X., {et~al.} 2016, \apjl, 823, L37

\bibitem[{{Ai} {et~al.}(2017){Ai}, {Fabian}, {Fan}, {Walker}, {Ghisellini},
  {Sbarrato}, {Dou}, {Wang}, {Wu}, \& {Feng}}]{ai2017}
{Ai}, Y., {Fabian}, A.~C., {Fan}, X., {et~al.} 2017, \mnras, 470, 1587

\bibitem[{{Arnaud}(1996)}]{1996ASPC..101...17A}
{Arnaud}, K.~A. 1996, in Astronomical Society of the Pacific Conference Series,
  Vol. 101, Astronomical Data Analysis Software and Systems V, ed. G.~H.
  {Jacoby} \& J.~{Barnes}, 17

\bibitem[{{Astropy Collaboration} {et~al.}(2018){Astropy Collaboration},
  {Price-Whelan}, {Sip{\H o}cz}, {G{\"u}nther}, {Lim}, {Crawford}, {Conseil},
  {Shupe}, {Craig}, {Dencheva}, {Ginsburg}, {VanderPlas}, {Bradley},
  {P{\'e}rez-Su{\'a}rez}, {de Val-Borro}, {Aldcroft}, {Cruz}, {Robitaille},
  {Tollerud}, {Ardelean}, {Babej}, {Bachetti}, {Bakanov}, {Bamford},
  {Barentsen}, {Barmby}, {Baumbach}, {Berry}, {Biscani}, {Boquien}, {Bostroem},
  {Bouma}, {Brammer}, {Bray}, {Breytenbach}, {Buddelmeijer}, {Burke},
  {Calderone}, {Cano Rodr{\'{\i}}guez}, {Cara}, {Cardoso}, {Cheedella},
  {Copin}, {Crichton}, {D{\'A}vella}, {Deil}, {Depagne}, {Dietrich}, {Donath},
  {Droettboom}, {Earl}, {Erben}, {Fabbro}, {Ferreira}, {Finethy}, {Fox},
  {Garrison}, {Gibbons}, {Goldstein}, {Gommers}, {Greco}, {Greenfield},
  {Groener}, {Grollier}, {Hagen}, {Hirst}, {Homeier}, {Horton}, {Hosseinzadeh},
  {Hu}, {Hunkeler}, {Ivezi{\'c}}, {Jain}, {Jenness}, {Kanarek}, {Kendrew},
  {Kern}, {Kerzendorf}, {Khvalko}, {King}, {Kirkby}, {Kulkarni}, {Kumar},
  {Lee}, {Lenz}, {Littlefair}, {Ma}, {Macleod}, {Mastropietro}, {McCully},
  {Montagnac}, {Morris}, {Mueller}, {Mumford}, {Muna}, {Murphy}, {Nelson},
  {Nguyen}, {Ninan}, {N{\"o}the}, {Ogaz}, {Oh}, {Parejko}, {Parley}, {Pascual},
  {Patil}, {Patil}, {Plunkett}, {Prochaska}, {Rastogi}, {Reddy Janga},
  {Sabater}, {Sakurikar}, {Seifert}, {Sherbert}, {Sherwood-Taylor}, {Shih},
  {Sick}, {Silbiger}, {Singanamalla}, {Singer}, {Sladen}, {Sooley},
  {Sornarajah}, {Streicher}, {Teuben}, {Thomas}, {Tremblay}, {Turner},
  {Terr{\'o}n}, {van Kerkwijk}, {de la Vega}, {Watkins}, {Weaver}, {Whitmore},
  {Woillez}, \& {Zabalza}}]{astropy2018}
{Astropy Collaboration}, {Price-Whelan}, A.~M., {Sip{\H o}cz}, B.~M., {et~al.}
  2018, ArXiv e-prints, arXiv:1801.02634

\bibitem[{{Ba{\~n}ados} {et~al.}(2015){Ba{\~n}ados}, {Venemans}, {Morganson},
  {Hodge}, {Decarli}, {Walter}, {Stern}, {Schlafly}, {Farina}, {Greiner},
  {Chambers}, {Fan}, {Rix}, {Burgett}, {Draper}, {Flewelling}, {Kaiser},
  {Metcalfe}, {Morgan}, {Tonry}, \& {Wainscoat}}]{banados2015a}
{Ba{\~n}ados}, E., {Venemans}, B.~P., {Morganson}, E., {et~al.} 2015, \apj,
  804, 118

\bibitem[{{Ba{\~n}ados} {et~al.}(2016){Ba{\~n}ados}, {Venemans}, {Decarli},
  {Farina}, {Mazzucchelli}, {Walter}, {Fan}, {Stern}, {Schlafly}, {Chambers},
  {Rix}, {Jiang}, {McGreer}, {Simcoe}, {Wang}, {Yang}, {Morganson}, {De Rosa},
  {Greiner}, {Balokovi{\'c}}, {Burgett}, {Cooper}, {Draper}, {Flewelling},
  {Hodapp}, {Jun}, {Kaiser}, {Kudritzki}, {Magnier}, {Metcalfe}, {Miller},
  {Schindler}, {Tonry}, {Wainscoat}, {Waters}, \& {Yang}}]{banados2016}
{Ba{\~n}ados}, E., {Venemans}, B.~P., {Decarli}, R., {et~al.} 2016, \apjs, 227,
  11

\bibitem[{{Ba{\~n}ados} {et~al.}(2018){Ba{\~n}ados}, {Venemans},
  {Mazzucchelli}, {Farina}, {Walter}, {Wang}, {Decarli}, {Stern}, {Fan},
  {Davies}, {Hennawi}, {Simcoe}, {Turner}, {Rix}, {Yang}, {Kelson}, {Rudie}, \&
  {Winters}}]{banados2018}
{Ba{\~n}ados}, E., {Venemans}, B.~P., {Mazzucchelli}, C., {et~al.} 2018, \nat,
  553, 473

\bibitem[{{Bischetti} {et~al.}(2017){Bischetti}, {Piconcelli}, {Vietri},
  {Bongiorno}, {Fiore}, {Sani}, {Marconi}, {Duras}, {Zappacosta}, {Brusa},
  {Comastri}, {Cresci}, {Feruglio}, {Giallongo}, {La Franca}, {Mainieri},
  {Mannucci}, {Martocchia}, {Ricci}, {Schneider}, {Testa}, \&
  {Vignali}}]{bischetti2017}
{Bischetti}, M., {Piconcelli}, E., {Vietri}, G., {et~al.} 2017, \aap, 598, A122

\bibitem[{{Brandt} {et~al.}(2002){Brandt}, {Schneider}, {Fan}, {Strauss},
  {Gunn}, {Richards}, {Anderson}, {Vanden Berk}, {Bahcall}, {Brinkmann},
  {Brunner}, {Chen}, {Hennessy}, {Lamb}, {Voges}, \& {York}}]{brandt2002}
{Brandt}, W.~N., {Schneider}, D.~P., {Fan}, X., {et~al.} 2002, \apjl, 569, L5

\bibitem[{{Brightman} {et~al.}(2013){Brightman}, {Silverman}, {Mainieri},
  {Ueda}, {Schramm}, {Matsuoka}, {Nagao}, {Steinhardt}, {Kartaltepe},
  {Sanders}, {Treister}, {Shemmer}, {Brandt}, {Brusa}, {Comastri}, {Ho},
  {Lanzuisi}, {Lusso}, {Nandra}, {Salvato}, {Zamorani}, {Akiyama}, {Alexander},
  {Bongiorno}, {Capak}, {Civano}, {Del Moro}, {Doi}, {Elvis}, {Hasinger},
  {Laird}, {Masters}, {Mignoli}, {Ohta}, {Schawinski}, \&
  {Taniguchi}}]{brightman2013}
{Brightman}, M., {Silverman}, J.~D., {Mainieri}, V., {et~al.} 2013, \mnras,
  433, 2485

\bibitem[{{Cash}(1979)}]{1979ApJ...228..939C}
{Cash}, W. 1979, \apj, 228, 939

\bibitem[{{Chen} {et~al.}(2017){Chen}, {Hickox}, {Goulding}, {Stern}, {Assef},
  {Kochanek}, {Brown}, {Harrison}, {Hainline}, {Alberts}, {Alexander},
  {Brodwin}, {Del Moro}, {Forman}, {Gorjian}, {Jones}, {Murray}, {Pope}, \&
  {Rovilos}}]{chen_c2017}
{Chen}, C.-T.~J., {Hickox}, R.~C., {Goulding}, A.~D., {et~al.} 2017, \apj, 837,
  145

\bibitem[{{Davies} {et~al.}(2018){Davies}, {Hennawi}, {Ba{\~n}ados}, {Simcoe},
  {Decarli}, {Fan}, {Farina}, {Mazzucchelli}, {Rix}, {Venemans}, {Walter},
  {Wang}, \& {Yang}}]{davies2018}
{Davies}, F.~B., {Hennawi}, J.~F., {Ba{\~n}ados}, E., {et~al.} 2018, ArXiv
  e-prints, arXiv:1801.07679

\bibitem[{{Duras} {et~al.}(2017){Duras}, {Bongiorno}, {Piconcelli}, {Bianchi},
  {Pappalardo}, {Valiante}, {Bischetti}, {Feruglio}, {Martocchia}, {Schneider},
  {Vietri}, {Vignali}, {Zappacosta}, {La Franca}, \& {Fiore}}]{duras2017}
{Duras}, F., {Bongiorno}, A., {Piconcelli}, E., {et~al.} 2017, \aap, 604, A67

\bibitem[{{Fabian}(2016)}]{fabian2016}
{Fabian}, A.~C. 2016, Astronomische Nachrichten, 337, 375

\bibitem[{{Fabian} {et~al.}(2014){Fabian}, {Walker}, {Celotti}, {Ghisellini},
  {Mocz}, {Blundell}, \& {McMahon}}]{fabian2014}
{Fabian}, A.~C., {Walker}, S.~A., {Celotti}, A., {et~al.} 2014, \mnras, 442,
  L81

\bibitem[{{Freeman} {et~al.}(2002){Freeman}, {Kashyap}, {Rosner}, \&
  {Lamb}}]{2002ApJS..138..185F}
{Freeman}, P.~E., {Kashyap}, V., {Rosner}, R., \& {Lamb}, D.~Q. 2002, \apjs,
  138, 185

\bibitem[{{Fruscione} {et~al.}(2006){Fruscione}, {McDowell}, {Allen},
  {Brickhouse}, {Burke}, {Davis}, {Durham}, {Elvis}, {Galle}, {Harris},
  {Huenemoerder}, {Houck}, {Ishibashi}, {Karovska}, {Nicastro}, {Noble},
  {Nowak}, {Primini}, {Siemiginowska}, {Smith}, \&
  {Wise}}]{2006SPIE.6270E..1VF}
{Fruscione}, A., {McDowell}, J.~C., {Allen}, G.~E., {et~al.} 2006, in
  \procspie, Vol. 6270, Society of Photo-Optical Instrumentation Engineers
  (SPIE) Conference Series, 62701V

\bibitem[{{Gallerani} {et~al.}(2017){Gallerani}, {Zappacosta}, {Orofino},
  {Piconcelli}, {Vignali}, {Ferrara}, {Maiolino}, {Fiore}, {Gilli},
  {Pallottini}, {Neri}, \& {Feruglio}}]{gallerani2017b}
{Gallerani}, S., {Zappacosta}, L., {Orofino}, M.~C., {et~al.} 2017, \mnras,
  467, 3590

\bibitem[{{Garmire} {et~al.}(2003){Garmire}, {Bautz}, {Ford}, {Nousek}, \&
  {Ricker}}]{2003SPIE.4851...28G}
{Garmire}, G.~P., {Bautz}, M.~W., {Ford}, P.~G., {Nousek}, J.~A., \& {Ricker},
  Jr., G.~R. 2003, in \procspie, Vol. 4851, X-Ray and Gamma-Ray Telescopes and
  Instruments for Astronomy., ed. J.~E. {Truemper} \& H.~D. {Tananbaum}, 28--44

\bibitem[{{Gehrels}(1986)}]{gehrels1986}
{Gehrels}, N. 1986, \apj, 303, 336

\bibitem[{{Just} {et~al.}(2007){Just}, {Brandt}, {Shemmer}, {Steffen},
  {Schneider}, {Chartas}, \& {Garmire}}]{just2007}
{Just}, D.~W., {Brandt}, W.~N., {Shemmer}, O., {et~al.} 2007, \apj, 665, 1004

\bibitem[{{Kalberla} {et~al.}(2005){Kalberla}, {Burton}, {Hartmann}, {Arnal},
  {Bajaja}, {Morras}, \& {P{\"o}ppel}}]{2005A&A...440..775K}
{Kalberla}, P.~M.~W., {Burton}, W.~B., {Hartmann}, D., {et~al.} 2005, \aap,
  440, 775

\bibitem[{{Lansbury} {et~al.}(2014){Lansbury}, {Alexander}, {Del Moro},
  {Gandhi}, {Assef}, {Stern}, {Aird}, {Ballantyne}, {Balokovi{\'c}}, {Bauer},
  {Boggs}, {Brandt}, {Christensen}, {Craig}, {Elvis}, {Grefenstette}, {Hailey},
  {Harrison}, {Hickox}, {Koss}, {LaMassa}, {Luo}, {Mullaney}, {Teng}, {Urry},
  \& {Zhang}}]{2014ApJ...785...17L}
{Lansbury}, G.~B., {Alexander}, D.~M., {Del Moro}, A., {et~al.} 2014, \apj,
  785, 17

\bibitem[{{Laor} {et~al.}(1994){Laor}, {Fiore}, {Elvis}, {Wilkes}, \&
  {McDowell}}]{laor1994}
{Laor}, A., {Fiore}, F., {Elvis}, M., {Wilkes}, B.~J., \& {McDowell}, J.~C.
  1994, \apj, 435, 611

\bibitem[{{Leipski} {et~al.}(2014){Leipski}, {Meisenheimer}, {Walter}, {Klaas},
  {Dannerbauer}, {De Rosa}, {Fan}, {Haas}, {Krause}, \& {Rix}}]{leipski2014}
{Leipski}, C., {Meisenheimer}, K., {Walter}, F., {et~al.} 2014, \apj, 785, 154

\bibitem[{{Luo} {et~al.}(2013){Luo}, {Brandt}, {Alexander}, {Harrison},
  {Stern}, {Bauer}, {Boggs}, {Christensen}, {Comastri}, {Craig}, {Fabian},
  {Farrah}, {Fiore}, {Fuerst}, {Grefenstette}, {Hailey}, {Hickox}, {Madsen},
  {Matt}, {Ogle}, {Risaliti}, {Saez}, {Teng}, {Walton}, \& {Zhang}}]{luo2013}
{Luo}, B., {Brandt}, W.~N., {Alexander}, D.~M., {et~al.} 2013, \apj, 772, 153

\bibitem[{{Luo} {et~al.}(2015){Luo}, {Brandt}, {Hall}, {Wu}, {Anderson},
  {Garmire}, {Gibson}, {Plotkin}, {Richards}, {Schneider}, {Shemmer}, \&
  {Shen}}]{luo2015}
{Luo}, B., {Brandt}, W.~N., {Hall}, P.~B., {et~al.} 2015, \apj, 805, 122

\bibitem[{{Lusso} \& {Risaliti}(2016)}]{lusso2016}
{Lusso}, E., \& {Risaliti}, G. 2016, \apj, 819, 154

\bibitem[{{Lusso} \& {Risaliti}(2017)}]{lusso2017}
---. 2017, \aap, 602, A79

\bibitem[{{Mahabal} {et~al.}(2005){Mahabal}, {Stern}, {Bogosavljevi{\'c}},
  {Djorgovski}, \& {Thompson}}]{mahabal2005}
{Mahabal}, A., {Stern}, D., {Bogosavljevi{\'c}}, M., {Djorgovski}, S.~G., \&
  {Thompson}, D. 2005, \apjl, 634, L9

\bibitem[{{Martocchia} {et~al.}(2017){Martocchia}, {Piconcelli}, {Zappacosta},
  {Duras}, {Vietri}, {Vignali}, {Bianchi}, {Bischetti}, {Bongiorno}, {Brusa},
  {Lanzuisi}, {Marconi}, {Mathur}, {Miniutti}, {Nicastro}, {Bruni}, \&
  {Fiore}}]{martocchia2017}
{Martocchia}, S., {Piconcelli}, E., {Zappacosta}, L., {et~al.} 2017, \aap, 608,
  A51

\bibitem[{{Mazzucchelli} {et~al.}(2017){Mazzucchelli}, {Ba{\~n}ados},
  {Venemans}, {Decarli}, {Farina}, {Walter}, {Eilers}, {Rix}, {Simcoe},
  {Stern}, {Fan}, {Schlafly}, {De Rosa}, {Hennawi}, {Chambers}, {Greiner},
  {Burgett}, {Draper}, {Kaiser}, {Kudritzki}, {Magnier}, {Metcalfe}, {Waters},
  \& {Wainscoat}}]{mazzucchelli2017b}
{Mazzucchelli}, C., {Ba{\~n}ados}, E., {Venemans}, B.~P., {et~al.} 2017, \apj,
  849, 91

\bibitem[{{Miniutti} \& {Fabian}(2004)}]{miniutti2004}
{Miniutti}, G., \& {Fabian}, A.~C. 2004, \mnras, 349, 1435

\bibitem[{{Moretti} {et~al.}(2014){Moretti}, {Ballo}, {Braito}, {Caccianiga},
  {Della Ceca}, {Gilli}, {Salvaterra}, {Severgnini}, \&
  {Vignali}}]{moretti2014}
{Moretti}, A., {Ballo}, L., {Braito}, V., {et~al.} 2014, \aap, 563, A46

\bibitem[{{Nanni} {et~al.}(2017){Nanni}, {Vignali}, {Gilli}, {Moretti}, \&
  {Brandt}}]{nanni2017}
{Nanni}, R., {Vignali}, C., {Gilli}, R., {Moretti}, A., \& {Brandt}, W.~N.
  2017, \aap, 603, A128

\bibitem[{{Nanni} {et~al.}(2018){Nanni}, {Gilli}, {Vignali}, {Mignoli},
  {Comastri}, {Vanzella}, {Zamorani}, {Calura}, {Lanzuisi}, {Brusa}, {Tozzi},
  {Iwasawa}, {Cappi}, {Vito}, {Balmaverde}, {Costa}, {Risaliti}, {Paolillo},
  {Prandoni}, {Liuzzo}, {Rosati}, {Chiaberge}, {Caminha}, {Sani}, {Cappelluti},
  \& {Norman}}]{nanni2018}
{Nanni}, R., {Gilli}, R., {Vignali}, C., {et~al.} 2018, ArXiv e-prints,
  arXiv:1802.05613

\bibitem[{{Page} {et~al.}(2014){Page}, {Simpson}, {Mortlock}, {Warren},
  {Hewett}, {Venemans}, \& {McMahon}}]{page2014}
{Page}, M.~J., {Simpson}, C., {Mortlock}, D.~J., {et~al.} 2014, \mnras, 440,
  L91

\bibitem[{{Park} {et~al.}(2006){Park}, {Kashyap}, {Siemiginowska}, {van Dyk},
  {Zezas}, {Heinke}, \& {Wargelin}}]{2006ApJ...652..610P}
{Park}, T., {Kashyap}, V.~L., {Siemiginowska}, A., {et~al.} 2006, \apj, 652,
  610

\bibitem[{{Richards} {et~al.}(2011){Richards}, {Kruczek}, {Gallagher}, {Hall},
  {Hewett}, {Leighly}, {Deo}, {Kratzer}, \& {Shen}}]{richards2011}
{Richards}, G.~T., {Kruczek}, N.~E., {Gallagher}, S.~C., {et~al.} 2011, \aj,
  141, 167

\bibitem[{{Risaliti} \& {Lusso}(2015)}]{risaliti2015}
{Risaliti}, G., \& {Lusso}, E. 2015, \apj, 815, 33

\bibitem[{{Salvaterra}(2015)}]{salvaterra2015}
{Salvaterra}, R. 2015, ArXiv e-prints, arXiv:1503.03072

\bibitem[{{Shemmer} {et~al.}(2006){Shemmer}, {Brandt}, {Schneider}, {Fan},
  {Strauss}, {Diamond-Stanic}, {Richards}, {Anderson}, {Gunn}, \&
  {Brinkmann}}]{shemmer2006}
{Shemmer}, O., {Brandt}, W.~N., {Schneider}, D.~P., {et~al.} 2006, \apj, 644,
  86

\bibitem[{{Stern}(2015)}]{stern2015}
{Stern}, D. 2015, \apj, 807, 129

\bibitem[{{Tanvir} {et~al.}(2009){Tanvir}, {Fox}, {Levan}, {Berger},
  {Wiersema}, {Fynbo}, {Cucchiara}, {Kr{\"u}hler}, {Gehrels}, {Bloom},
  {Greiner}, {Evans}, {Rol}, {Olivares}, {Hjorth}, {Jakobsson}, {Farihi},
  {Willingale}, {Starling}, {Cenko}, {Perley}, {Maund}, {Duke}, {Wijers},
  {Adamson}, {Allan}, {Bremer}, {Burrows}, {Castro-Tirado}, {Cavanagh}, {de
  Ugarte Postigo}, {Dopita}, {Fatkhullin}, {Fruchter}, {Foley}, {Gorosabel},
  {Kennea}, {Kerr}, {Klose}, {Krimm}, {Komarova}, {Kulkarni}, {Moskvitin},
  {Mundell}, {Naylor}, {Page}, {Penprase}, {Perri}, {Podsiadlowski}, {Roth},
  {Rutledge}, {Sakamoto}, {Schady}, {Schmidt}, {Soderberg}, {Sollerman},
  {Stephens}, {Stratta}, {Ukwatta}, {Watson}, {Westra}, {Wold}, \&
  {Wolf}}]{tanvir2009}
{Tanvir}, N.~R., {Fox}, D.~B., {Levan}, A.~J., {et~al.} 2009, \nat, 461, 1254

\bibitem[{{Venemans} {et~al.}(2017){Venemans}, {Walter}, {Decarli},
  {Ba{\~n}ados}, {Carilli}, {Winters}, {Schuster}, {da Cunha}, {Fan}, {Farina},
  {Mazzucchelli}, {Rix}, \& {Weiss}}]{venemans2017c}
{Venemans}, B.~P., {Walter}, F., {Decarli}, R., {et~al.} 2017, \apjl, 851, L8

\bibitem[{{Vietri} {et~al.}(2018){Vietri}, {Piconcelli}, {Bischetti}, {Duras},
  {Martocchia}, {Bongiorno}, {Marconi}, {Zappacosta}, {Bisogni}, {Bruni},
  {Brusa}, {Comastri}, {Cresci}, {Feruglio}, {Giallongo}, {La Franca},
  {Mainieri}, {Mannucci}, {Ricci}, {Sani}, {Testa}, {Tombesi}, {Vignali}, \&
  {Fiore}}]{vietri2018}
{Vietri}, G., {Piconcelli}, E., {Bischetti}, M., {et~al.} 2018, ArXiv e-prints,
  arXiv:1802.03423

\bibitem[{{Vito} {et~al.}(2018){Vito}, {Brandt}, {Stern}, {Assef}, {Chen},
  {Brightman}, {Comastri}, {Eisenhardt}, {Garmire}, {Hickox}, {Lansbury},
  {Tsai}, {Walton}, \& {Wu}}]{vito2018b}
{Vito}, F., {Brandt}, W.~N., {Stern}, D., {et~al.} 2018, \mnras, 474, 4528

\bibitem[{{Volonteri}(2012)}]{volonteri2012a}
{Volonteri}, M. 2012, Science, 337, 544

\bibitem[{{Wachter} {et~al.}(1979){Wachter}, {Leach}, \&
  {Kellogg}}]{1979ApJ...230..274W}
{Wachter}, K., {Leach}, R., \& {Kellogg}, E. 1979, \apj, 230, 274

\bibitem[{{Wang} {et~al.}(2016){Wang}, {Wu}, {Fan}, {Yang}, {Yi}, {Bian},
  {McGreer}, {Yang}, {Ai}, {Dong}, {Zuo}, {Jiang}, {Green}, {Wang}, {Cai},
  {Wang}, \& {Yue}}]{wang-feige2016}
{Wang}, F., {Wu}, X.-B., {Fan}, X., {et~al.} 2016, \apj, 819, 24

\end{thebibliography}



\end{document}